\documentstyle[11pt,paspconf,epsf]{article}

\begin{document}

\title{Study of the TeV Emission from Mkn~501 with the Stereoscopic
Cherenkov Telescope System of HEGRA}
\author{H. Krawczynski for the HEGRA Collaboration}
\affil{ MPI f\"ur Kernphysik, Postfach 10 39 80, D-69029 Heidelberg (Germany)}
\begin{abstract}
The HEGRA system of 4 Imaging Atmospheric Cherenkov Telescopes (IACTs)
has been used since March 1997 for a comprehensive study of the 
gamma-ray emission from the BL Lac object Mkn~501 in the energy 
range above 500~GeV.
Taking advantage of the unique capabilities of the IACT system, i.e.\ 
an unprecedented flux sensitivity in the TeV energy range 
and an unique energy resolution of better than 20\% for individual 
TeV photons, detailed information about
the temporal and spectral characteristics of the source during its 
spectacular bright phase in 1997 is reported. 
Further more, using the large HEGRA and {\it Rossi X-Ray Timing Explorer} 
All Sky Monitor (RXTE ASM) data bases,  the correlation of the X-ray 
and the TeV activity of the source is discussed.
Finally, an outlook over present and future activities is given.
\end{abstract}
\section{Introduction}
In the first two years after its discovery as a TeV $\gamma$-ray source,
the BL Lac object Mkn~501 showed fluxes well below the persistent flux  
of the Crab Nebula (Quinn et al.\ 1996, Bradbury et al. 1997).
In 1997 the source went into a state of surprisingly high 
activity and dramatic variability, outshining during several nights
the brightest known source in the TeV sky, the Crab Nebula, 
by factors as large as $\sim10$. 
In this paper we report on detailed studies of
this spectacular bright phase performed with the HEGRA IACT system.
The IACT system (Daum et al.\ 1997) is located on the Roque de los Muchachos on the 
Canary Island of La Palma, (lat.\ 28.8$^\circ$ N,long.\ 17.9$^\circ$, 2200 m a.s.l.). 
It is formed by 5, during 1997 by 4, identical IACTs - 
one at the center and 4 (during 1997, 3) at the corners of a 100~m by 100~m square area.
Each telescope is equipped with a segmented 8.5~m$^2$ mirror 
and a 4.3$^\circ$ field of view high resolution camera consisting 
of 271 pixels of 0.25$^\circ$ diameter. 
Exploiting the stereoscopic observation technique (simultaneous observation of air 
showers under widely differing viewing angles with two or more 
Cherenkov telescopes, see Aharonian et al.\ 1997) 
the system achieves a low energy threshold of 500\,GeV, 
an excellent angular resolution of 0.1$^\circ$, 
an energy resolution of better than 20\% (all for individual photons), 
and a flux sensitivity $\nu F_\nu$ at 1\, TeV of
$10^{-11} \, \rm ergs/cm^2 sec$  $\simeq$\,1/4 Crab for 1 hour 
of observation time (S/$\sqrt{\rm B}$=5$\sigma$ with a system of 4 IACTs).
The 4 IACT system started operation in fall 1996.
\section{Data sample and analysis method}
The analysis of this paper is based on 110 hours of Mkn~501 data
acquired between March 16th, 1997 and October 1st, 1997 
under optimal weather conditions, with the optimal detector performance, 
and with Mkn~501 being more than 45$^\circ$ above the horizon.
Altogether about $38,000$ Mkn 501 photons were recorded, making it possible to
verify the source location with an accuracy of 35 arcsec (P\"uhlhofer et al.\ 1997).
Since the IACT system provides an unprecedented signal to noise ratio, 
loose $\gamma$/hadron-separation cuts can be used to extract 
the Mkn~501 signal and to suppress the background of charged cosmic rays
which accept a large fraction of $\sim80\%$ of the 
$\gamma$-rays at all energies above 1~TeV.
By this means, the systematic uncertainties associated with
uncertainties in energy dependent cut efficiencies are minimized.
The analysis, i.e.\ the cut optimization and the calculation of effective detection areas
and cut efficiencies, is based on detailed Monte Carlo simulations 
which have been checked experimentally using cosmic ray data (hadron induced showers) 
and Mkn~501 and Crab data (photon induced showers). 
A more detailed description of the analysis tools and also
of the temporal characteristics of Mkn 501 can be found in 
(Aharonian et al.\ 1998).
\section{Time-averaged 1997 Mkn 501 energy spectrum}
\begin{figure}[t]
\plottwo
{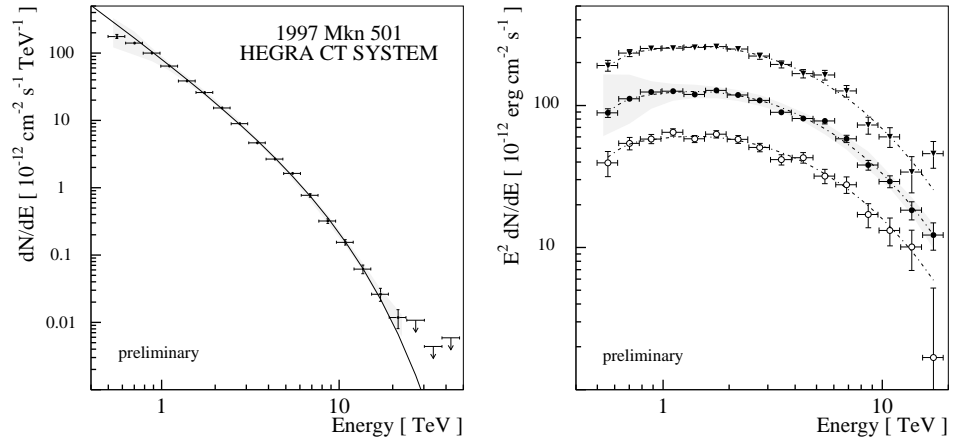}
{krawcz_lett11.epsi}
\vspace*{-0.2cm}
\caption{On the left side, the time-averaged 1997 Mkn 501 energy spectrum is shown.
The right side shows the time-averaged 1997 Mkn 501 SED (middle) and the 
time-averaged SED for all nights with a differential flux at 2 TeV below 
1.6 (lower SED) and above 3 (upper SED) times 10$^{-11}$ 
${\rm cm^{-2} s^{-1} TeV^{-1}}$.
The dotted-dashed line shows the shape of the average SED
overlaid over all three SEDs. 
Both sides: vertical error bars show the statistical errors, 
the shaded region the systematic error on the shape of the spectrum
(see text).
}
\vspace*{-0.2cm}
\end{figure}
The time-averaged Mkn 501 energy spectrum is shown in Fig.\ 1 (left side)
over the energy region from 500 GeV to 25 TeV.
For determining the spectrum down to energies below 800 GeV,
the analysis is restricted to the 80\,h of low energy threshold data 
taken with Mkn 501 at altitudes $>$~60$^\circ$.
The systematic error on the absolute energy scale is 15\%. 
The shaded region shows our current conservative 
estimate of the additional systematic error on the shape of the spectrum. 
It is mainly caused by uncertainties in the effective areas
near detection threshold.
The error bars in vertical direction show the statistical errors and
the error bars in horizontal direction indicate the energy resolution 
of the IACT System. The spectrum is smooth over the whole energy range and it is clearly curved. 
Although the exact shape of the spectrum above 10\,TeV is still preliminary, 
the Mkn 501 emission clearly extends into the energy range well above 10\,TeV.
A $\chi^2$-analysis yields a 2\,$\sigma$ lower limit on the minimum photon 
energy of the signal of 18\,TeV. 
A fit of the data over the energy region of small systematic errors,
i.e.\ from 1.25 TeV to 50 TeV, with a power law model with an exponential 
cut off gives:
\[
dN/dE \, = \, 9.7\, 
\pm 0.3\, ({\rm stat})\, \pm 2.0\, ({\rm syst}) \cdot 10^{-11}\,
E^{-1.9 \, \pm 0.06 \,({\rm stat})\, \pm 0.07 \,({\rm syst})}\vspace*{-0.3cm}
\]
\[ 
\hspace*{1.5cm}
\exp\left[-E/(5.7 \,\pm  1.1 ({\rm stat}) \,\pm 0.6 \,({\rm syst})\, 
{\rm TeV})
\right]
{\rm cm^{-2} s^{-1} TeV^{-1}}.
\]
In Fig.\ 1 (right side) the spectral energy distribution (SED) $\nu F_{\nu}$ is shown
for the mean spectrum and for all days with a differential flux
at 2 TeV below 1.6 and above 3 times 10$^{-11}$ 
${\rm cm^{-2} s^{-1} TeV^{-1}}$.
Seemingly, the SEDs peak in the energy range between 500 GeV and 2 TeV, although,
due to the systematic uncertainties, a peak in the energy range below 500 GeV
can not be excluded.
The three SEDs have within the statistics the same shape.
Thus we do not find any evidence for a correlation of the emission strength at 2 TeV 
and the spectral shape in the energy range from 500 GeV to 20 TeV. 
Most interestingly, as described further below, the time-averaged spectrum also fits the 
diurnal energy spectra statistically satisfactorily. 
The observed spectral shape is invariant during the whole 1997 observation period.
\section{The temporal characteristics of the 1997 Mkn 501 emission}
The stereoscopic IACT system makes it possible to determine differential TeV spectra even on
diurnal basis. 
\begin{figure}[t]
\plotone
{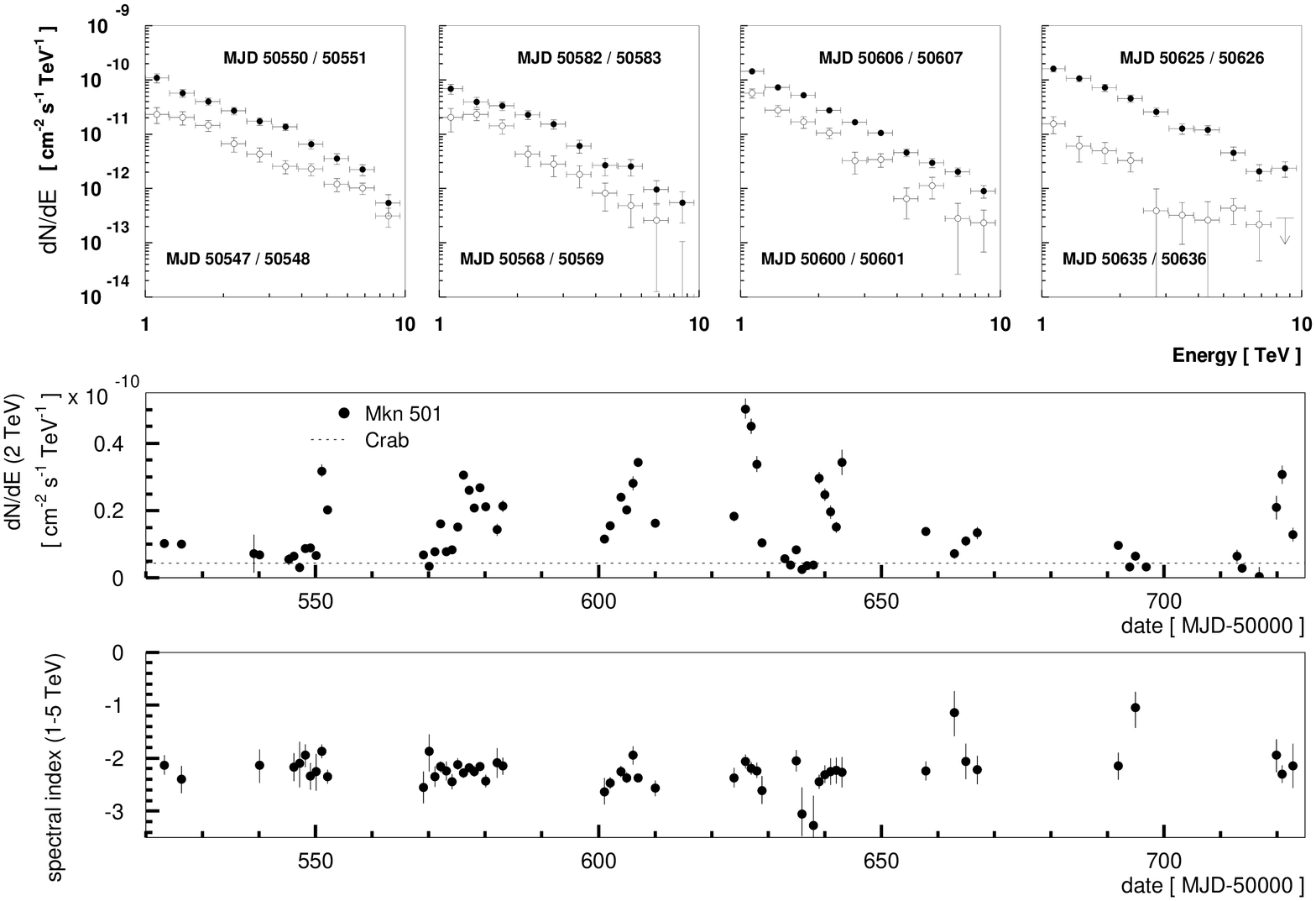}
\vspace*{-0.2cm}
\caption{The upper panel shows the $\gamma$-ray spectra of eight individual nights.
For each of the four data periods March/April, April/May, May/June, and June/July, 1997
a night of weak emission and a night of strong emission has
been chosen (upper limit has 2$\sigma$ confidence level).
The two lower panels show the
diurnal diff.\ fluxes at 2~TeV and spectral indices (1-5\,TeV) for the 1997 Mkn~501 data. 
Spectral indices are shown only for the days with 
sufficient statistics, i.e.\ with errors on the spectral index $<$0.5.
Measurement gaps are due to bad weather or shining moon.
All three plots: only statistical errors, see text for systematic errors.
(MJD 50550 is April 12th, 1997)
}
\vspace*{-0.2cm}
\end{figure}
Figure 2 (upper panel) shows the differential spectra obtained for 8 exemplary individual nights
in the energy range from 1 to 10 TeV. 
We do not find any diurnal spectrum with a shape which deviates significantly from the
time-averaged 1997 spectrum. The temporal evolution of emission intensity and spectral 
steepness have been studied by fitting power law models to the diurnal spectra 
in the energy range from 1 to 5~TeV. In Fig.\ 2 (2 lower panels) the results are shown.
As before, the error bars show the statistical errors only. The systematic error on the
flux amplitude deriving from the 15\% uncertainty in the energy scale is approximately
20\% and the systematic uncertainty on the spectral index is 0.1.
The emission intensity, i.e.\ the differential flux at 2~TeV, 
varies dramatically from a fraction of a Crab unit to $\sim10$ Crab units,
the peak emission being recorded on MJD 50625/50626.
In contrast, the differential spectral indices from 1 to 5~TeV are rather stable.
Only two $3\,\sigma$-deviations from the mean value -2.25 have been found, namely 
for the night MJD 50550/50551 the spectral index is $-$1.87 +0.13 $-$0.14 and
for the night MJD 50694/50695 it is $-$1.05 +0.30 $-$0.38.

A dedicated search for the shortest time scales of flux variability has been carried out.
The time gradient of the flux computed with adjacent diurnal flux amplitudes $\Delta t$ hours apart 
corresponds to shortest increase/decay times $\tau=\Delta t/\Delta \ln({\rm flux})$
in the order of 15\,h.
Variability within individual nights could only marginally be detected for the two nights
MJD 50576/50577 and MJD 50606/50607. The trial corrected chance probability for
more significant variability is 0.4\% and 1\% for both nights respectively.
The flux variabilities of these two days 
correspond to increase/decay times in the order of 5 h.
\section{Correlation X-ray~/~TeV}
\begin{figure}[t]
\plotone{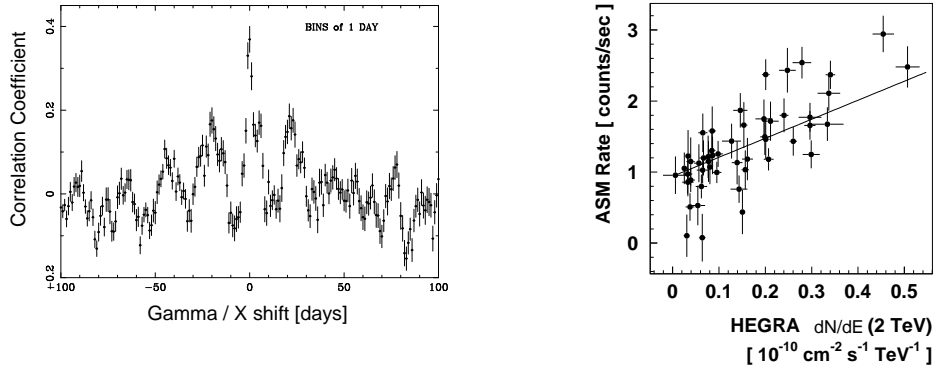}
\vspace*{-0.6cm}
\vspace*{-1ex}
\caption{The left side shows the correlation coefficient of the 
daily HEGRA fluxes at 2\,TeV
and the one-day ASM count rates (2-12\,keV)
as a function of time shift $\Delta\,t$ between the 
two fluxes ($\Delta\,t>0$ means TeV precedes X-ray).
The right side shows the correlation of the daily fluxes at 2\,TeV
with the one-day ASM count rates for $\Delta t\,=\,0$. 
Superimposed is a straight
line fit to the data.}
\vspace*{-0.2cm}
\end{figure}
The RXTE ASM (Remillard \& Levine 1997) data have been used
to study the correlation between the 2 to 12~keV and the TeV emission intensities.
Figure 3 (left side) shows the Discrete Correlation Function DCF (Edelson \& Krolik 1988) 
as function of the time lag $\Delta t$ between X-ray and TeV variability, 
as computed from the HEGRA diurnal flux amplitudes at 2~TeV and the ASM 2-12~keV
count rate, the latter for each day averaged over all measurements within a 24~h
interval centered close to 0:00 UTC.
The DCF shows evidence for a weak correlation between X-ray and TeV activity with a time
lag between X-ray and TeV emission smaller than or equal to one day.
The DCF computed for $\Delta t \,=\,0$ with 50 pairs of data is 0.37$\pm$0.03.
Due to the limited number of $\simeq$50 pairs of data entering the
determination of the DCF, the significance of the correlation is modest.
Depending on the assumptions about the autocorrelation properties of the X-ray and TeV 
emission, the chance probability for larger DCF values is computed to lie between
0.43\% and 8\%.
In Fig.\ 3 (right side), the correlation between
the diurnal X-ray and TeV fluxes is shown for $\Delta t \,=\,0$.
The straight line fit to the data indicates a much larger relative flux variability
in the TeV energy range than in the 2 to 12 keV energy band.
\section{Outlook}
The 1997 high emission phase of Mkn 501 made it possible to 
study this BL Lac object in the TeV energy range with unprecedented
signal to noise ratio during a long time period of more than 6 months. 
The IACT system of HEGRA has been used to obtain a wealth
of detailed spectral and temporal information.
Most interestingly, within the statistical accuracy, the shape of the energy spectrum is 
constant during the whole observation period and extends well into the
energy range above 10\,TeV.
A deep understanding of the spectral properties is rendered difficult
since several effects combine to give the observed spectrum, e.g.\  
the spectrum of the emitting electrons, the spectrum of possible
Inverse Compton seed photons, internal $\gamma_{\rm TeV},\gamma_{\rm O,UV}$ absorption of the TeV photons
in the source, and intergalactic absorption of the TeV photons 
in $\gamma_{\rm TeV}\,\gamma_{\rm IR}\,\rightarrow \,e^+\,e^-$ processes
by the Diffuse Extragalactic Background Radiation (DEBRA).
Note that already the pure fact of the registration of TeV photons with energies exceeding 10 TeV
yields a sensitive upper limit on the largely unconstrained DEBRA in the wavelength region
from 1 to 20 microns.
Due to very general arguments concerning the emitted $\gamma$-ray luminosity,
the optical depth $\tau$ of the DEBRA for TeV photons cannot exceed 1 by much 
more than one order of magnitude. 
The condition $\tau\,<\tau_0 \simeq 10$ yields for the DEBRA density
$n(\varepsilon)$ at energy $\varepsilon$ the upper limit
$\varepsilon^2\,n(\varepsilon)/({\rm 10^{-3} eV/cm^3})\,<\,(\tau_0/5)\,({\rm H}_0/ {\rm (60 km/s Mpc)})\,/\,(\varepsilon/{\rm eV})$
where H$_{0}$ is the Hubble constant, with only small corrections depending on the shape of the spectrum.

Models will further be constrained 
by intensive {\it multiwavelength campaigns}
and by studying {\it more sources}.
The analysis of the Mkn 501 and Mkn 421 multiwavelength campaigns 
performed during 1997 and 1998 with participation of the HEGRA IACT 
array is underway. Further more, the IACT system has extensively 
been used to search for new TeV emitting BL Lac sources,
although, up to now, without positive evidence.
Members of the HEGRA collaboration pursue the installation of two next generation
IACT installations aiming at a sensitivity increase by one order of magnitude. 
HESS, a stereoscopic system of at first 4, in the second stage 16 IACTs of the 10 m
diameter class for $\gamma$-ray astronomy at energies above 40 GeV (Hofmann 1997)
will probably start operation in the year 2001. 
MAGIC will be a dedicated ``low energy threshold'' stand alone IACT for 
$\gamma$-ray observations above an energy threshold of 10 GeV (Lorenz 1997).
\acknowledgments
We thank the Instituto de Astrof\'{\i}sica de Canarias (IAC) for
supplying excellent working conditions at La Palma. 
HEGRA is supported by the BMBF (Germany) and CYCIT (Spain).
The RXTE ASM data has been obtained through the High Energy Astrophysics
Science Archive Research Center Online Service, provided by the 
NASA/\-Goddard Space Flight Center.


\begin{references}
\reference {Aharonian F.A., Hofmann W., Konopelko A.K., V\"olk H.J.\ 
1997, Astropart. Phys.\ 6, 343}
\reference {Aharonian F.A., Akhperjanian A.G., Barrio J.A., et al.,  
1998, A\&A accepted for publication, astro-ph/9808296}
\reference {Bradbury S.M., Deckers T., Petry D., et al., 1997, A\&A 320, L5}
\reference {Daum A., Hermann G., He{\ss} M., et al., 1997, Astropart.\ Phys.\ 8, 1} 
\reference {Edelson R.A., Krolik J.H., 1988, ApJ 333, 646}
\reference Hofmann, W., 1997. In: 
Proc. Towards a Major Atmospheric Cherenkov Detector-V, ed. O.C. de Jager, 405
\reference Lorenz, E.C., 1997. In: Proc. 25th ICRC, Durban, 5, 177
\reference {P\"uhlhofer G.,  Daum A., Hermann G., et al., 1997, Astropart. Phys.\ 8, 101}
\reference {Quinn J., Akerlof C.W., Biller S., et al., 1996, ApJ 456, L83}
\reference {Remillard R.A., Levine M.L., 1997, astro-ph/9707338}
\end{references}
\end{document}